\documentclass[aps,prl,twocolumn,reprint,superscriptaddress]{revtex4-2}
\usepackage{graphicx,color}
\usepackage{amsmath}
\usepackage{amssymb}
\usepackage{colordvi}
\usepackage{mathrsfs}
\usepackage{bm}
\usepackage{verbatim}
\usepackage{dcolumn}
\usepackage{epsfig}
\usepackage{subfigure}

\usepackage{mathtools}
\usepackage{braket}

\usepackage{hyperref}

\newcommand{\di}{\mathrm i}

\begin{document}
\title{Anisotropic Gyromagnetic Ratio and Orthogonal Einstein-de Haas Effect}

\author{Rui Xue}
\affiliation{International Centre for Quantum Design of Functional Materials, CAS Key Laboratory of Strongly-Coupled Quantum Matter Physics, and Department of Physics, University of Science and Technology of China, Hefei, Anhui 230026, China}
\affiliation{Hefei National Laboratory, University of Science and Technology of China, Hefei 230088, China}
\author{Zhenhua Qiao}
\email[Correspondence author:~]{qiao@ustc.edu.cn}
\affiliation{International Centre for Quantum Design of Functional Materials, CAS Key Laboratory of Strongly-Coupled Quantum Matter Physics, and Department of Physics, University of Science and Technology of China, Hefei, Anhui 230026, China}
\affiliation{Hefei National Laboratory, University of Science and Technology of China, Hefei 230088, China}
\author{Yang Gao}
\email[Correspondence author:~]{ygao87@ustc.edu.cn}
\affiliation{International Centre for Quantum Design of Functional Materials, CAS Key Laboratory of Strongly-Coupled Quantum Matter Physics, and Department of Physics, University of Science and Technology of China, Hefei, Anhui 230026, China}
\affiliation{Hefei National Laboratory, University of Science and Technology of China, Hefei 230088, China}
\author{Qian Niu}
\affiliation{International Centre for Quantum Design of Functional Materials, CAS Key Laboratory of Strongly-Coupled Quantum Matter Physics, and Department of Physics, University of Science and Technology of China, Hefei, Anhui 230026, China}

\date{\today{}}

\begin{abstract}
  {We theoretically demonstrate an orthogonal Einstein-de Haas effect, where the rotation of ferromagnetic materials is caused by the change of magnetization in the direction orthogonal to the rotation axis. This amounts to an anisotropic gyromagnetic ratio. To reveal its microscopic origin, we treat the spin-orbit coupling as a perturbation, integrate out the electronic degree of freedom, and show that in collinear ferromagnets the phonon angular momentum admits a dipolar structure in the spin-order space due to the constraint of the spin group symmetry. The spin-flipping and spin-conserving parts of the spin-orbit coupling contribute differently to such a dipolar structure. All these features are exemplified in a lattice electron-phonon model with ferromagnetic order and $C_{1h}$ point group symmetry. Our work lays the ground for revealing the connection between phonon angular momentum and general spin-order configurations.}
\end{abstract}
\maketitle

%[Noted]-Phonon chirality has attracted considerable research interest. In systems where time-reversal symmetry is broken, a nonzero phonon angular momentum arise, and chiral phonons couple with the magnetic field, contributing to the phonon magnetic moment. In previous studies, the phonon magnetic moment has been attributed to different microscopic mechanisms: the Born effective charge and electron-phonon coupling.

\begin{figure}[t]
	\includegraphics[width=7cm,angle=0]{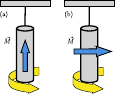}
	\caption{Schemes of (a) Conventional~(parallel) Einstein-de Haas effect, and (b) Orthogonal Einstein-de Haas effect.}
	\label{fig_fig1}
\end{figure}

Chiral phonons, i.e., phonons that carry nontrivial angular momentum, have attracted great interest in recent years~\cite{Zhang2014,Zhang2015,Garanin2015,Zhu2018,Zhang2018,Miao2018,Nomura2019,Zhang2020,Li2021,Zhang2022,Juraschek2022,Zabalo2022,Zhang2023,Ueda2023,Gao2023,Bonini2023,Tang2024,Mustafa2025,Xue2025,Zhang2025,He2020,Cheng2020,Saito2019,Ren2021,Hu2021,Bistoni2021,Saparov2022,Lujan2024,Che2025}. The angular momentum associated with the chiral motion can be transferred to other degrees of freedom such as electron and magnon~\cite{Chen2020,He2020,Lujan2024,Saito2019,Ren2021,Hu2021,Bistoni2021,Saparov2022,Zhang2019,Wang2024,Che2025}, supporting rich chiral response phenonema~\cite{Strohm2005,Sheng2006,Inyushkin2007,Kagan2008,Wang2009,Zhang2010,Juraschek2022}.
In magnetic materials, the fundamental parameter governing the angular momentum is the cross gyromagnetic ratio. It linearly relates the phonon angular momentum to the spin order and hence directly leads to the Einstein-de Haas effect as shown in Fig.~\ref{fig_fig1}(a)~\cite{Zhang2014,Garanin2015}. It is often assumed that such gyromagnetic ratio is a material-dependent but numerical scalar, rendering the angular momentum to be parallel with the magnetization.

Microscopically speaking, under the Born-Oppenheimer approximation, the fast electronic motion affects the slow atomic motion in two ways, i.e., the energy landscape and the gauge potential. The latter one leads to the molecular Berry curvature of electrons~\cite{Saito2019,Cheng2020,Ren2021,Hu2021,Bistoni2021,Saparov2022,Bonini2023}, which affects the dynamic structure of the atomic motion, producing the circular motion and the resulting angular momentum~\cite{Zhang2014,Zhang2015}. As an electronic property, the molecular Berry curvature is subject to full group symmetry of the original crystals. This then determines the gyromagnetic ratio and hence the precise dependence of the phonon angular momentum on magnetic order. Since such structure is derived, there is not a priori reason for the phonon contribution to the gyromagnetic ratio to be a numerical scalar.

In this work, we systematically explore the dependence of the phonon angular momentum on spin order. By treating the spin-orbit coupling as perturbation and applying the constraint of spin group symmetry, we find that the phonon angular momentum exhibits a dipolar structure in the spin order space, leading to a tensorial gyromagnetic ratio. Such anisotropic gyromagnetic ratio leads to an orthogonal Einstein-de Haas effect where the rotation of the materials is caused by the change of magnetization on the orthogonal direction, as shown in Fig. ~\ref{fig_fig1}(b). We have also provided detailed guidance for searching material candidates. Moreover, the spin-flipping and spin-conserving parts of the spin-orbit coupling have different contributions to this effect. These features are exemplified in detail in a cubic lattice with $C_{1h}$ point-group symmetry. Our work exhibits rich dipolar structure in the fundamental dependence of the phonon angular momentum on spin order.

\textit{Dipolar structure of phonon angular momentum.---} We focus on the phonon angular momentum in collinear ferromagnetic crystals and establish the theory of the crossed gyromagnetic ratio. We consider the following Hamiltonian describing both electrons and phonons:
\begin{align}\label{eq_fullH}
\hat{H}=&\frac{\bm p_e^2}{2m}+U(\bm r_e, \{\bm R_i\})+ J(\bm r,\{\bm R_i\})\hat{\bm m}\cdot \bm \sigma +\frac{\hbar}{m^2c^2} \hat{\bm O}\cdot \bm \sigma\notag\\
&+ \sum_i\frac{\bm P_i^2}{2M_i}+U(\{\bm R_i\})\,,
\end{align}
where the first four terms describe the electron's Hamiltonian including kinetic energy, periodic potential, exchange potential, and spin orbit coupling. $\hat{\bm m}$ points out the direction of ferromagnetic spin order,  $\{\bm R_i\}$ represents atomic coordinates, and $\hat{\bm O}=\bm \nabla_{\bm r_e} U(\bm r_e, \{\bm R_i\})\times \hat{\bm p}_e$. The last two terms describe the atomic motion, including kinetic energy and potential energy. We emphasize that we aim to calculate the angular momentum of phonons~\cite{Zhang2014}, i.e.,
\begin{align}\label{eq_lph}
\bm L_{ph}=\sum_i\langle \bm u_i\times\dot{ \bm u}_i\rangle\,,
\end{align}
where $\bm u_i$ is the displacement field of the i-th atom, and the average is with respect to the Bose-Einstein distribution of phonons.

Two comments are in order for $\hat{H}$. First, by applying the Born-Oppenheimer approximation and integrating out the electronic degree of freedom, the atomic potential energy $U(\{\bm R_i\})$ and the atomic momentum $\bm P_i$ are shifted by an effective scalar and vector potential~\cite{Saito2019,Cheng2020,Ren2021,Saparov2022}, respectively. Second, it is known that the spin-lattice coupling is essential for phonon chirality. In paramagnetic materials it can come from the Barnett effect~\cite{Barnett1915,Barnett1935,Ono2015}. However, in ferromagnetic materials, it mainly comes from the molecular Berry curvature of electrons~\cite{Ren2021,Saparov2022}, which is the curl of vector potential.

Obviously, a net phonon angular momentum requires breaking the time-reversal symmetry. In Eq.~\eqref{eq_fullH}, the only time-reversal symmetry breaking term is the exchange coupling. It is then straightforward to prove the Onsager's reciprocal relation of $\bm L_{ph}(\bm M)=-\bm L_{ph}(-\bm M)$. In the phonon Hamiltonian, by integrating out the electronic motion, $\bm M$ manifests itself from the molecular Berry curvature $\bm \Omega_i$~\cite{Saparov2022}, i.e.,
\begin{align}
\hat{H}^\prime=\bm \Omega_i \cdot (\bm u_i\times \dot{\bm u}_i)\,,
\end{align}
where $\bm \Omega_i=\nabla_{\bm u_i}\times \bm A_{\bm u_i}$ with $\bm A_{\bm u_i}=i\langle \psi|\partial_{\bm u_i}|\psi\rangle$ with $|\psi\rangle$ being the mean-field wave function when the $i$-th lattice site has a displacement $\bm u_i$. Since $\bm \Omega_i$ flips sign under the time-reversal operation, one can then have the spin-lattice coupling and finally the phonon angular momentum. Here we assume the on-site spin-lattice coupling.

However, the time-reversal symmetry breaking is insufficient for the phonon angular momentum. The spin-group symmetry breaking by the spin-orbit coupling is equally important. For collinear ferromagnets, if the full Hamiltonian in Eq.~\eqref{eq_fullH} does not have the spin-orbit coupling, it respects the spin group symmetry which always contains a $C_{2x}^s T$ element~\cite{GosalbezMartinez2015}, where $C_{2x}^s$ only acts on spin with its rotation axis perpendicular to $ \hat{\bm m}$. This necessarily dictates $\bm L_{ph}$ to vanish. This result can hold even with coplanar spin order, for either ferromagnetic or antiferromagnetic case.

Compared with the exchange coupling associated with the spin order, the spin-orbit coupling can be reasonably treated as perturbation in deriving the phonon angular momentum. It is noted that, in the electronic part of the Hamiltonian, the energy scale of the spin-orbit coupling is on the order of 10 to 100 meV~\cite{Dunn1961,Herman1964,Stoehr2006,Tanaka2008,Naito2010,Yuan2017,Stamokostas2018,Khomskii2020}. Although it is much smaller than the hopping energy and exchange energy that are on the order of eV, it is on the same order of atomic force for phonons~\cite{Kittel2013,Saito1998}. However, the spin-orbit coupling does not directly act on phonons. Its effect can be obtained by integrating out the electronic degree of freedom, and is at least one order smaller than the effect of other electronic energy scales. Therefore, to explain $\bm \Omega_i$ and the resulting $\bm L_{ph}$, one can treat the spin-orbit coupling as a perturbation.

To reveal the role of the spin-orbit coupling, we adopt the framework in Ref.~\cite{Liu2025} and define a separate spin frame whose $z$-axis always follows $\hat{\bm m}$. By writing the orbital degree of freedom in the lab frame and the spin degree of freedom in the spin frame, we find that the first three terms have fixed forms even when $\hat{\bm m}$ rotates in space~($\hat{\bm m}\cdot \bm \sigma\rightarrow \sigma_z$). Since the spin-orbit coupling changes, then a set of spin-orbit vectors $\bm \ell^a$ is needed:
\begin{align}
\frac{\hbar}{m^2c^2} \hat{\bm O}\cdot \bm \sigma\rightarrow \frac{\hbar}{m^2c^2} \hat{ O}_j \ell_j^a  \sigma_a\,.
\end{align}

Since the spin-orbit coupling is linear in $\bm \ell^a$, the expansion for both $\bm \Omega_i$ and $\bm L_{ph}$ can reduce to a power series of $\bm \ell^a$. By focusing on $\bm L_{ph}$ and applying the constraint of spin-only group symmetry, we find that~\cite{suppl}
\begin{align}
\label{eq_cons1} \bm L_{ph}(\bm \ell^x,\bm \ell^y,\bm \ell^z)&=-\bm L_{ph}(\bm \ell^x,-\bm \ell^y,-\bm \ell^z)\,,\\
\label{eq_cons2} \bm L_{ph}(\bm \ell^x,\bm \ell^y,\bm \ell^z)&=\bm L_{ph}(\bm \ell^x(\phi), \bm \ell^y(\phi), \bm \ell^z)\,,
\end{align}
where $\bm \ell^x(\phi)=\bm \ell^x \cos\phi+\bm \ell^y \sin\phi$ and $\bm \ell^y(\phi)=-\bm \ell^x \sin\phi+\bm \ell^y \cos\phi$ for arbitrary $\phi$. By using the connection between $\ell_i^a$ and the orientation of ferromagnetic spin order, we find that the resulting phonon angular momentum should adopt similar power series~\cite{suppl}, i.e., 
\begin{align}\label{eq_multi}
(L_{ph})_i=p_{ij} \hat{m}_j+o_{ijk\ell}\hat{m}_j \hat{m}_k \hat{m}_\ell+\cdots.
\end{align}
Such a multipolar expansion only has odd-order poles, in consistent with the requirement of Onsager's relation due to the time-reversal symmetry. Moreover, as the order of poles increases, the magnitude decreases as higher order poles are at higher order of the spin-orbit coupling.

We then focus on the leading order contribution, which is due to the dipolar structure of $\bm L_{ph}$ in the spin-order space. The inverse of the dipole $p_{ij}$ directly gives rise to the cross gyromagnetic ratio $\gamma_{ij}$, i.e.,
\begin{align}
\gamma_{ij}=|\bm M|(p^{-1})_{ij}\,,
\end{align}
where $|\bm M|$ is the strength of spin order.

As a rank-2 tensor, the dipole $p_{ij}$ can be decomposed into a trace part, a symmetric part, and an antisymmetric part as follows:
\begin{align}
p_{ij}=p_0 \delta_{ij}+p_{ij}^s+\epsilon_{ijk}d_k\,,
\end{align}
where $\epsilon_{ijk}$ is the levi-Civita symbol. These three parts have distinct effects on the gyromagnetic ratio and hence on the phonon angular momentum. The trace part alone locks the direction of $\bm L_{ph}$ to be always parallel with $ \hat{\bm m}$, so that the gyromagnetic ratio is just a numerical scalar, as often used previously. In sharp contrast, the antisymmetric part $\bm d$ gives rise to a phonon angular momentum that is always perpendicular to $\hat{\bm m}$. The symmetric part $p_{ij}^s$ has both effects. It can be diagonalized by choosing appropriate principal axises. When $\hat{\bm m}$ is along any principal axis, the contribution to $\bm L_{ph}$ from $p_{ij}^s$ is parallel with $\hat{\bm m}$; otherwise, it deviates from $\hat{\bm m}$ but is usually not perpendicular to $\hat{\bm m}$.

These features have a striking effect in the Einstein-de Haas experiment. The total angular momentum of the system is conserved and can be expressed as follows~\cite{Zhang2014}:
\begin{align}
	\bm L_{tot} = \bm L_{lat} + \bm L_{ph} + \bm L_{spin} + \bm L_{orb}\,.
\end{align} 
The trace part $p_0$ corresponds to the conventional Einstein-de Haas effect, where the flipping of $\hat{\bm m}$ causes the crystal to rotate about $\hat{\bm m}$, as shown in Fig.~\ref{fig_fig1}(a). However, the antisymmetric part of $p_{ij}$ indicates that $\bm L_{ph}$ has a component that is perpendicular to $\hat{\bm m}$. As $\hat{\bm m}$ flips, so is this perpendicular component of $\bm L_{ph}$. This suggests an orthogonal Einstein-de Haas effect, as shown in Fig.~\ref{fig_fig1}(b). This effect can also be induced by $p_{ij}^s$, as long as $\hat{\bm m}$ is not along any principal axis.

We now analyze the symmetry requirement for the orthogonal Einstein-de Haas effect. In collinear ferromagnets, the spin group is the direct product of the crystallographic group and the spin-only group~\cite{Liu2022,Chen2025}. As a result, the dipole $p_{ij}$ is only constrained by the crystallographic point group. For cubic group, both $p_{ij}^s$ and $\bm d$ vanish, meaning that the orthogonal Einstein-de Haas effect generally exists. With a nonzero $p_{ij}^s$, one only needs to make $\hat{\bm m}$ deviate from any principal axis of $p_{ij}^s$. Meanwhile, $\bm d$ can exist in three cases: 1) only one rotational axis exists, and $\bm d$ is along this axis; 2) only one mirror plane exists, and $\bm d$ is perpendicular to the mirror plane; 3) both rotation axis and mirror plane exist, with the rotation axis being perpendicular to the mirror plane. This symmetry analysis is summarized in Supplementary Materials. Cobalt can be considered a concrete example, where both the rotation axis and magnetizaiton direction are oblique against the $c$-axis.

\begin{figure}[t]
	\includegraphics[width=9cm,angle=0]{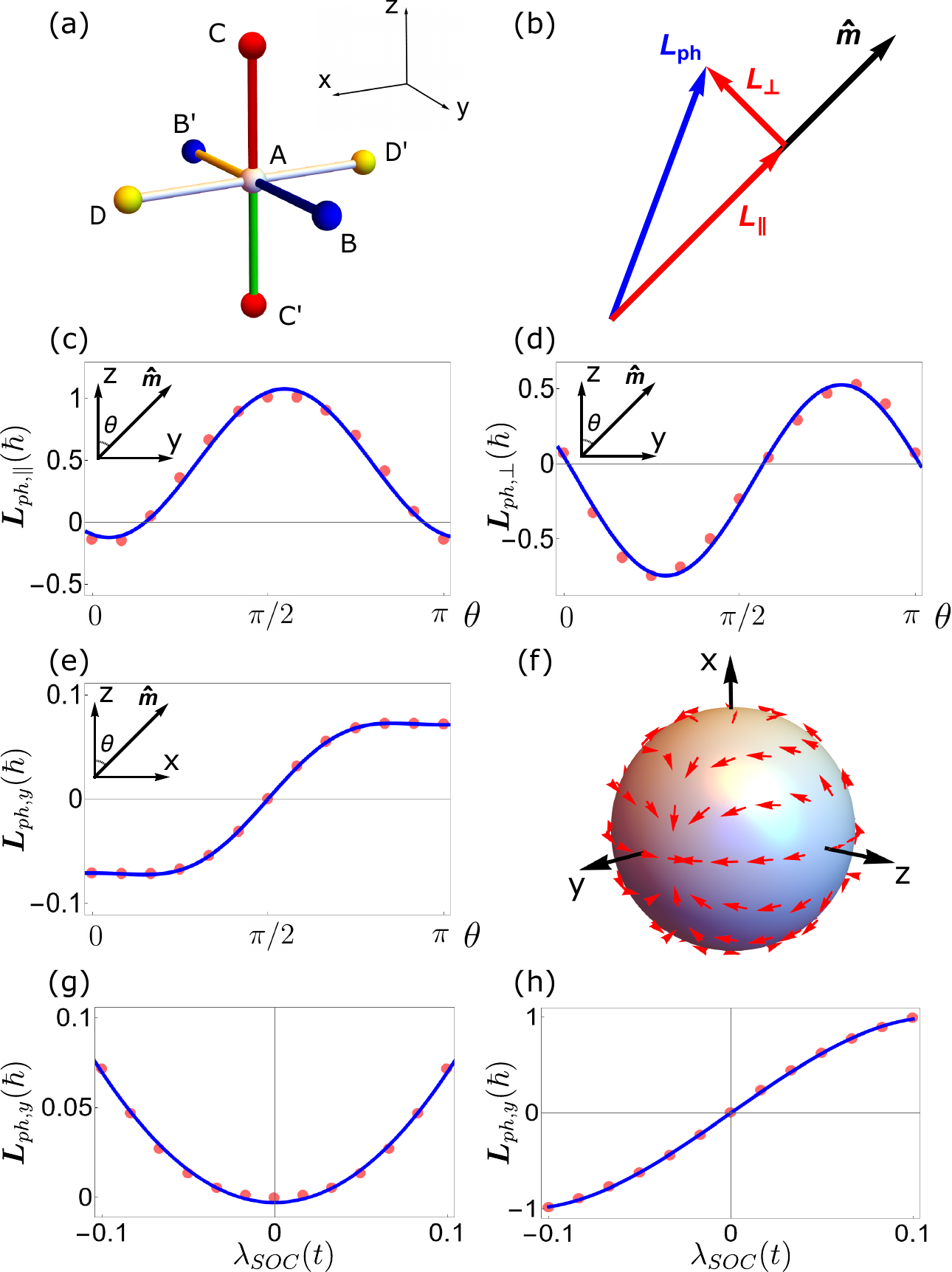}
	\caption{The phonon angular momentum of the lattice model. (a) The unit cell of the lattice model. Bonds with different hopping amplitudes are showing with different colors. (b) The decomposition of phonon angular momentum into parallel and perpendicular components. (c) and (d) The $L_{ph,\parallel}$ and $L_{ph,\perp}$ with $\hat{\bm m}$ rotating in the $zy$ plane. (e) The $L_{ph,y}$ with $\hat{\bm m}$ rotating in the $xz$ plane. (f) The dipolar structure of the $L_{ph,\perp}$. (g) and (h) The  $L_{ph,\perp}$ as a function of the strength of the spin-orbit coupling with $\hat{\bm m}\parallel \hat{z}$ and $\bm \delta$ in the spin-orbit coupling along $\hat{z}$~(in (g)) and $\hat{y}$ direction~(in (h)). }
	\label{fig_fig2}
\end{figure}

\textit{Lattice Model}.---As a concrete example, we consider a crystal with $C_{1h}$ point group symmetry. The electronic Hamiltonian is written as:
\begin{align}
\hat{H}_{el} =& \sum_{\braket{i,j},\alpha} t_{ij} c_{i\alpha}^{\dagger} c_{j\alpha} + \sum_{i,\alpha,\beta} J \hat{m} \cdot \vec{\sigma}_{\alpha \beta} c_{i\alpha}^{\dagger} c_{i \beta} \notag\\
&+ \sum_{\langle i,j\rangle,\alpha,\beta}\di \lambda \, \vec{\delta}\cdot (\hat{R}_{ij}\times\vec{\sigma}_{\alpha\beta})c_{i\alpha}^{\dagger} c_{j \beta} +\mathrm{h.c.},
\end{align}
where the first term is the nearest-neighbour hopping term, the second term is the exchange coupling and the last term is the Rashba spin-orbit coupling. The hopping parameters $t_{ij}$ are chosen in a way to be consistent with the $C_{1h}$ point-group symmetry with the mirror plane perpendicular to $\hat{x}$ direction. Using this electronic Hamiltonian, one can calculate the molecular Berry curvature to obtain the Raman coupling. The phonon Hamiltonian is written as:
\begin{align}
\hat{H}_{ph}=\frac{1}{2}\sum_{a}\bm p_{a}^{2} + \frac{1}{2}\sum_{ab}(u_a)_iD_{ij}^{ab}(u_b)_j+\sum_{a}\bm \Omega_a\cdot (\bm u_a\times \dot{\bm u}_a)]\,,
\end{align}
where the force constant $D_{ij}^{ab}$ is consistent with $C_{1h}$ point group symmetry and $\bm \Omega_a$ is calculated using $\hat{H}_{el}$. System details are provided in Supplementary Materials~\cite{suppl}.

We then study the phonon angular momentum. The angular momentum can be decomposed into components that are respectively orthogonal to and parallel with $\hat{\bm m}$ (i.e., $L_{ph,\parallel}$ and $L_{ph,\perp}$), as shown in Fig.~\ref{fig_fig2}(b). The $C_{1h}$ group symmetry dictates that
\begin{align}
	\bm p = &
	\begin{pmatrix}
		p_{xx}^{s} & 0 & 0  \\
		0 & p_{yy}^{s} & p_{yz}^{s} + d_{x} \\
		0 & p_{yz}^{s} - d_{x} & p_{zz}^{s}
	\end{pmatrix} \,.
\end{align}
When the spin order rotates in the $yz$ plane, $\bm L_{ph}$ lies within the same plane. In specific, we have
\begin{align}\label{eq_dipstrzy}
	L_{ph,\parallel} =& \,\frac{p_{yy}^{s}+p_{zz}^{s}}{2}-\frac{p_{yy}^{s}-p_{zz}^{s}}{2}\cos(2\theta)+p_{yz}^{s}\sin(2\theta), \notag\\
	L_{ph,\perp} =& -d_x+\frac{1}{2}(p_{zz}^{s}-p_{yy}^{s})\sin(2\theta)-p_{yz}^s \cos(2\theta)\,.
\end{align}
Since $L_{ph,\perp}$ directly corresponds to the orthogonal phonon angular momentum, we find that as the magnetic order rotates within the $xz$ plane, the orthogonal Einstein-de Haas effect should generally exist. Moreover, $L_{ph,\perp}$ originates from both antisymmetric and anisotropic symmetric parts of the dipole, and hence is on the same order of magnitude with $L_{ph,\parallel}$, which arises from the symmetric part. In other words, the orthogonal Einstein-de Haas effect can be very large. The numerical results of $L_{ph,\parallel}$ and $L_{ph,\perp}$ are shown in Figs.~\ref{fig_fig2}(c) and \ref{fig_fig2}(d). One can see that the angle dependence agrees well with the prediction in Eq.~\eqref{eq_dipstrzy}, and that $L_{ph,\perp}$ is comparable with $L_{ph,\parallel}$, both of which confirm the validity of our theory.

When the spin order rotates in the $xz$ plane, the orthogonal component of $\bm L_{ph}$ has two parts. The first one lies within $xz$ plane and originates from the anisotropic symmetric dipole. Similar to Eq.~\eqref{eq_dipstrzy}, it has a period of $\pi$. In comparison, the second one is perpendicular to $xz$ plane and reads as
\begin{align}\label{eq_dipperpzx}
	L_{ph,y}=& (d_x+p_{yz}^s) \cos\theta\,.
\end{align}
Both the antisymmetric and anisotropic symmetric dipoles contribute, and the resulting phonon angular momentum has a period of $2\pi$. This angular dependence is confirmed by our numerical results, as shown in Fig.~\ref{fig_fig2}(e).

To offer a general guideline for observing orthogonal Einstein-de Haas effect, it is helpful to exhibit the distribution of $\bm L_{ph,\perp}$ with all possible spin order orientations, as plotted in Fig.~\ref{fig_fig2}(f). One can find that the anisotropic symmetric part of the dipole contributes to the dipolar structure of $\bm L_{ph,\perp}$ while the antisymmtric part contributes to the circular structure in $yz$ plane. Experimentally, one usually rotates the spin order within a fixed plane. One can then find the intersection of such a plane at the spheretical surface, which then shows the pattern of phonon angular momentum. 

Besides the angular dependence, our theory also illuminates on the roles of different parts of the spin-orbit coupling. There are two types of spin-orbit coupling, i.e., spin-flipping part and spin conserving part, which contain the parts of spin operator that is perpendicular to and parallel with the spin order orientation. For anomalous Hall effect that requires breaking time-reversal symmetry, it is important to identify the role of these two types of spin-orbit coupling. It is thus informative to know their roles in phonon angular momentum. 

We first discuss the role of spin-flipping part of the spin-orbit coupling. We fix both the spin order and the vector $\bm \delta$ in the spin-orbit coupling to be along the $z$ direction. The spin-orbit coupling is expressed as $H_{SOC} = \di \lambda \, [(\hat{R}_{ij})_{x} (\sigma_{y})_{\alpha\beta} - (\hat{R}_{ij})_{y} (\sigma_{x})_{\alpha\beta}]c_{i\alpha}^{\dagger} c_{j \beta}$, which only contains the spin-flipping part of spin order coupling. Specifically, we have $\bm \ell^x=(1,0,0)$, $\bm \ell^y=(0,1,0)$, and $\bm \ell^z=0$. According to the constraints in Eqs.~\eqref{eq_cons1} and \eqref{eq_cons2}, the leading order contribution should be written as~\cite{suppl}
\begin{align}
L_{ph,i}=\beta_{ijk}(\ell_j^x\ell_k^y-\ell_k^x \ell_j^y)\,.
\end{align}
In other words, both the parallel and perpendicular parts of the phonon angular momentum should be at least second order with respect to the spin-orbit coupling strength. This can be confirmed by our numerical results in Fig.~\ref{fig_fig2}(g).

We then discuss the role of the spin-conserving part of the spin-orbit coupling. We still fix the spin order to be along $z$ direction but now the vector $\bm \delta$ in the spin-orbit coupling is along $y$ direction. The spin-orbit coupling becomes $H_{SOC} = \di \lambda \, [(\hat{R}_{ij})_{z} (\sigma_{x})_{\alpha\beta} - (\hat{R}_{ij})_{x} (\sigma_{z})_{\alpha\beta}]c_{i\alpha}^{\dagger} c_{j \beta}$, which contains both spin-flipping and spin-conserving parts of the spin order coupling.  Specifically, $\bm \ell^x=(1,0,0)$, $\bm \ell^y=0$, and $\bm \ell^z=(0,0,1)$. Since a nonzero second order contribution involves both $\bm \ell^x$ and $\bm \ell^y$, the leading two order contributions to $\bm L_{ph}$ should be the linear and cubic order, as shown in Fig.~\ref{fig_fig2}(h).

The cubic order contribution may support a special orthogonal Einstein-de Haas effect. This can be easily observed in ferromagnetic materials with cubic point group. In this case, the dipole $p_{ij}$ reduces to a scalar and hence cannot contribute to the orthogonal phonon angular momentum. However, the nonlinearity from the octupolar order can still contribute. This then greatly expands candidate materials hosting orthogonal Einstein-de Haas effect. For example, it may exist in common ferromagnets when the spin order is not along high-symmetry directions.

Our theory can be extended to nonmagnetic materials, where the phonon angular momentum is induced by external magnetic field~\cite{suppl}. In this case, the magnetic field enters the electron Hamiltonian through the Peierls substitution and can be treated as a perturbation. The resulting molecular Berry curvature and phonon angular momentum can be expanded in power series of the magnetic field. The phonon angular momentum should show similar angular dependence on the direction of the magnetic field but its dependence on spin-orbit coupling is completely different. Our theory can also be used to analyze the coupling between phonon angular momentum and more complicated spin textures, such as non-collinear and non-coplanar magnets. The expansion over spin-orbit coupling is general. The only difference is that the spin group structure can be more complicated than that in collinear ferromagnets.

In summary, we propose an orthogonal Einstein-de Haas effect where the phonon angular momentum can have a perpendicular component orthogonal to the magnetization. It originates from the dipolar structure in the orientation space of the spin order. Our theory also reveals distinct roles of different parts of the spin-orbit coupling in generating phonon angular momentum.

\begin{acknowledgements}
This work was financially supported by the National Key R\&D Program of China (Grant Nos. 2024YFA1408103), National Natural Science Foundation of China (12234017, 12374164, 12474158, and 12488101), Innovation Program for Quantum Science and Technology (2021ZD0302800). We also thank the Supercomputing Center of University of Science and Technology of China for providing high-performance computing resources.
\end{acknowledgements}

\bibliographystyle{apsrev4-2}
\bibliography{mainbblOEdH}

\end{document}